\documentclass[a4paper,11pt]{article}
\usepackage{jheppub}

\title{Charged Galileon black holes}

\author[a]{Eugeny Babichev,}
\author[a]{Christos Charmousis,}
\author[b]{and Mokhtar Hassaine}

\affiliation[a]{Laboratoire de Physique Th\'eorique (LPT), Univ.
Paris-Sud, CNRS UMR 8627, F-91405 Orsay, France}
\affiliation[b]{Instituto de Matem\'atica y F\'{\i}sica, Universidad de Talca,
Casilla 747, Talca, Chile.}
\emailAdd{eugeny.babichev@th.u-psud.fr}
\emailAdd{christos.charmousis@th.u-psud.fr}
\emailAdd{hassaine@inst-mat.utalca.cl}

\abstract{ We consider an Abelian gauge field coupled to a particular truncation of Horndeski theory. The Galileon field has translation symmetry and couples non minimally both to the metric and the gauge field. When the gauge-scalar coupling is zero the gauge field reduces to a standard Maxwell field. By taking into account the symmetries of the action, we construct charged black hole solutions. Allowing the scalar field to softly break symmetries of spacetime we construct black holes where the scalar field is regular on the black hole event horizon. Some of these solutions can be interpreted as the equivalent of Reissner-Nordstrom black holes of scalar tensor theories with a non trivial scalar field. A self tuning black hole solution found previously is extended to the presence of dyonic  charge without affecting whatsoever the self tuning of a large positive cosmological constant. Finally, for a general shift invariant scalar tensor theory we demonstrate that the scalar field Ansatz and method we employ are mathematically compatible with the field equations. This opens up the possibility for novel searches of hairy black holes in a far more general setting of Horndeski theory.}

\begin{document}

\maketitle

\section{Introduction}

The stationary state of a black hole in General Relativity can be determined by a limited number of physical charges which are conserved quantities subject to a Gauss law:  mass, electric and magnetic
charge and angular momentum. Quite generically any matter sector, which one tries to
attach to a black hole is either eaten up and
hidden behind the horizon, or/and expelled away from the black hole. After relaxation of the system, when we return to the stationary
state again, the global parameters of the black hole may have changed, but "hair" does not survive. All properties of a black hole are determined from asymptotic measurements. This statement is known as the no-hair paradigm for black holes \cite{wheeler}.
Not surprisingly, it is impossible to formulate a general no-hair theorem for arbitrary matter{\footnote{When one includes a cosmological constant in the action the presence of an extra length scale makes no hair theorems far weaker (see for example \cite{Martinez:2004nb}). In particular, no hair theorems generically do not work for asymptotically adS spacetimes. This is due to the trapping nature of adS spacetime.}}.
One needs to make certain precise assumptions about  properties of
matter and how this latter is coupled to gravity. Using these assumptions it is possible to
formulate specific no-hair theorems only in some particular cases. For example, in the case of a general
minimally coupled scalar field (when the Lagrangian of the scalar field is a
function of the canonical kinetic term and the field itself), under
reasonable assumptions including rather weak ones for the scalar potential, one can prove that scalar hair is absent
for a static black hole, (see \cite{Bekenstein:1998aw} for an overview). For a stationary black hole one has to assume stronger conditions for the scalar potential but even they do not cover all possibilities. Indeed very recently a loop-hole has been found and a hairy stationary black hole has been constructed numerically by Herdeiro and Radu \cite{Herdeiro:2014goa, Herdeiro:2015gia}. This is achieved for a (minimally coupled) massive complex scalar field. The crucial property is that the scalar field does not have the same symmetries as those imposed on space-time. This is an allowed anzatz for the field equations as long as the energy-momentum tensor of the scalar field {\it does} have the spacetime symmetries (while the scalar does not). As such a black hole is numerically constructed taking advantage of the superradiant instability of stationary black holes. This method fails for static spacetimes where a no hair theorem can be also shown to hold \cite{Pena:1997cy}, \cite{Graham:2014ina} even though the scalar field is allowed to break spacetime symmetries.

The situation is more complex for non-minimally coupled scalar fields, in
particular, when the theories under consideration are allowed to be of higher order but with second order field equations. This is the case of  Horndeski/Galileon
theory~\cite{Horndeski,deffayet}. Due to the higher order nature of the theory, additional branches of solutions appear yielding sometimes to surprising results. Furthermore, the galileon as it is called in such theories, in
general kinetically mixes with the graviton, therefore this opens
other possibilities to evade the no-hair theorems. An elegant approach in order to elaborate a no hair theorem for shift-symmetric
Galileons has been put forward in~\cite{Hui:2012qt}, where the use
of shift symmetry allowed  (again, with some assumptions) to prove such a result.

One can, however, look for possibilities to allow for the presence of scalar hair by slightly twisting or bifurcating some of the hypotheses. In the case of Galileons, a non-trivial possibility in order to evade the result of Ref.~\cite{Hui:2012qt} would be
to relax the assumption of staticity of the Galileon field while assuming it to be true for spacetime.
Indeed, for a standard canonical field with a potential
the staticity assumption is natural, since one expects that at  spatial
infinity the scalar evolves to its
minimum (as it occurs during cosmological expansion). In the case of a shift-symmetric Galileon, however, the
static ansatz for the scalar field is not natural anymore, due to
the non-trivial dynamics of the Galileon. Indeed, depending on the
choice of the Lagrangian, the Galileon field may have several
attractors for the cosmological dynamics. The non-trivial attractors
--- in particular, with de-Sitter metric --- have a non-trivial
evolution of the scalar field and the scalar ends up depending
linearly on cosmic time.

The cosmological evolution of the scalar field poses the boundary
conditions for the galileon far from the black hole. Therefore a static ansatz for the black hole galileon would clearly be at odds with a cosmological galileon,
since at large radii the assumption is clearly violated. Thus a
natural choice for a shift-symmetric Galileon theory would be, in
fact, to assume that the scalar field is time-dependent. The
simplest way to implement this assumption is to take a linear time
dependence of the scalar field\footnote{Due to the non-trivial
mixing of the scalar and the graviton, interesting physical
phenomena emerge in local physics, when the scalar field is
time-dependent~\cite{Babichev:2010kj}.}. This assumption obviously
bifurcates the no-go theorem of Ref.~\cite{Hui:2012qt}, giving a
principal possibility to create a non-trivial configuration of the
scalar field.

On the other hand, one can look for a loophole in the proof of a
no-go theorem. For a shift-symmetric Galileon, an important step of
the proof relies on the physical requirement, that the current
associated with the shift-symmetry, is non-diverging at the horizon.
This means that the radial component of the current, $J^r$, must be
zero. The non-trivial step in Horndeski theories is to then prove that $J^r=0$ leads indeed to a
trivial configuration of the scalar field~\cite{Hui:2012qt}. This is not trivial and in some cases as we will see simply not true since the theory at hand is of higher order and other possibilities do exist.

In Ref.~\cite{Babichev:2013cya} both loopholes have been
investigated to find black hole solutions with non-trivial scalar
field (see also \cite{Sotiriou:2013qea} for a complimentary discussion). A simple model has been considered: the standard kinetic term
plus a Galileon term $\sim G_{\mu\nu}\nabla^\mu\phi\nabla^\nu\phi$.
A combination of these terms allows for the $r$-component of the
current to have zero value, while at the same time the scalar field
is not trivial. Moreover, the scalar field
itself has been found to be regular at the horizon (which is a much
stronger condition than the finiteness of the norm of the current),
for the time-dependent configurations\footnote{A number of works
studied similar theories, finding solutions with a static scalar
field, \cite{Rinaldi:2012vy,Minamitsuji:2013ura,Cisterna:2014nua}.
However, unlike the solution of \cite{Babichev:2013cya}, the scalar
field explodes at the horizon for these solutions. In
\cite{Korolev:2014hwa} the authors constructed wormhole solutions
with the same type of the Lagrangian.}. The same strategy later has
been applied to find a number of solutions in other Galileon-type
models, see
e.g.~\cite{Bravo-Gaete:2013dca,Kobayashi:2014eva,Charmousis:2014zaa}.

In this paper we extend the work~\cite{Babichev:2013cya} to
include a gauge field. In Sec.~\ref{Seceoms} we write down the full
action and its equations of motion. Then we assume a specific
spherically symmetric ansatz. This ansatz is static for the
metric and the gauge field, but time-dependent for the scalar field. In Sec.~\ref{Secsol}
the  system of differential equations is completely integrated,
leading to two algebraic equations. This system of algebraic
equations fully describes the solution, though in an implicit form.
To find explicit and interesting solutions we consider judicious choices of
parameters and integration constants, for which we are able to write
down the solutions analytically. In the last section we take a step back and discuss the generality of our scalar field Anzatz and most importantly its mathematical consistency. In other words: is our scalar (time-dependent) anzatz just a way to obtain particular very special solutions for special theories or is there a deeper and more general reasoning behind the integrability properties? This, we believe is a crucial and central question in the further study of Galileon black holes. To answer this question, we consider a general scalar tensor theory with translational invariance. We assume a static and spherically symmetric spacetime along with a linear time dependent (plus radial) scalar field anzatz.  We then prove that the equations of motion are not overdetermined and the anzatz is in fact a consistent one leading always to an "integrable system". Most importantly we spotlight an intrinsic property of the
system of equations that pinpoints the $J_r=0$ property of the scalar field.  The time-radial component of the metric equations of motion (the flux of the scalar)
 is actually proportional to the radial component of the Noether
current $J_r$, associated with the shift-symmetry. Thus setting $J_r=0$ (with a non trivial scalar) allows the immediate resolution of 2 field equations rendering the system not only consistent but more tractable and in certain cases completely integrable analytically. We finally discuss the obtained results in Sec.~\ref{SecDisc}.

\section{Field equations and Ansatz}
\label{Seceoms}

In this Section we shall discuss the choice of the action,
the field equations, as well as  the ansatz we will opt for. The
action we consider includes the metric, one real scalar field and
one Abelian gauge field. The latter part of the action which is new in this paper is constructed in two steps,
described in detail below. We consider a static and spherically symmetric ansatz for the spacetime metric as well as the gauge field.  We will crucially  consider an additional (to the radial) linear time dependence for the scalar field which will be primordial for the regularity of the black hole solutions we will find. The consistency of this ansatz will be discussed in the final
part of the paper, since it is true for a general Galileon/gauge theory admitting translational invariance.

\subsection{Construction of the action}
We consider a four-dimensional action, which depends on a metric,
one scalar field and one $U(1)$ gauge field. The action we will take into account
can be thought of as a two-step construction, following the scheme,
first presented in \cite{Babichev:2013cya}. Let us start from the
standard action, containing the Einstein-Hilbert term, a cosmological
constant and the standard Maxwell gauge field,
\begin{equation}\label{protoaction}
    S_0 =
    \int\sqrt{-g}\, d^{4}x\   \left[ R - 2\,\Lambda -\frac{1}{4}\,F_{\mu \nu}\,F^{\mu \nu}\right],
\end{equation}
where $F_{\mu\nu} \equiv \partial_\mu A_\nu - \partial_\nu A_\mu$ is
the field strength. The variation of each of the term in the above
action with respect to the metric gives correspondingly (up to
constants), the Einstein tensor $G_{\mu\nu}$, the metric
$g_{\mu\nu}$, and the energy-momentum tensor of the Maxwell field,
\begin{eqnarray}\label{TE}
T_{\mu \nu}^{(M)}\equiv \frac{1}{2}\left( F_{\mu \sigma}
F_{\nu}^{\phantom{\nu} \sigma}-\frac{1}{4}\,g_{\mu \nu}\,F_{\alpha
\beta} F^{\alpha\beta}\right).
\end{eqnarray}

We build the scalar interaction terms in the action by contracting
the result of the variation of (\ref{protoaction}) with the tensor
$\nabla^\mu\phi \nabla^\nu\phi$, made out of the
derivatives of the scalar field. With this procedure, we obtain,
\begin{equation}
    R \to G_{\mu\nu} \nabla^\mu\phi \nabla^\nu\phi, \;\;
    \Lambda\to (\partial\phi)^2, \;\;
     F_{\mu \nu}\,F^{\mu \nu} \to T_{\mu \nu}^{(M)}\nabla^\mu\phi \nabla^\nu\phi.
\end{equation}
Now, allowing arbitrary coefficients in front of the obtained terms,
the full action which we study in this paper reads
\begin{equation}\label{action1}
\begin{aligned}
S[g_{\mu\nu},\phi,A_{\mu}] =& \\
\int\sqrt{-g}d^{4}x\  & \left[ R  -2\,\Lambda -\frac{1}{4}\,F_{\mu
\nu}\,F^{\mu\nu} + \beta\,
G_{\mu\nu}\nabla^{\mu}\phi\nabla^{\nu}\phi -\eta\,(\partial
\phi)^{2} -\gamma \,T_{\mu \nu}\,
\nabla^{\mu}\phi\nabla^{\nu}\phi\right],
\end{aligned}
\end{equation}
where $\eta$, $\beta$, $\gamma$ and $\Lambda$ are constants and
$T_{\mu \nu}^{(M)}$ is given by (\ref{TE}). We omitted the overall
$M_P^2$ (where $M_P$ is the Planck mass). Since we are only
interested in classical solutions, this overall constant does not
enter the final result. In our conventions the scalar field is
dimensionless, as well as the gauge field. With these definitions,
$\eta$ is a dimensionless constant, $\beta$ and $\gamma$ are
constants with the dimension [{\it length}]$^2$ and the dimension of
$\Lambda$ is [{\it mass}]$^2$. The case $\beta=0$, $\gamma=0$ and
$\eta=1/2$ corresponds to the canonical minimally coupled scalar
field, and minimally coupled Maxwell electromagnetic field. The term $\beta\, G_{\mu\nu}\nabla^{\mu}\phi\nabla^{\nu}\phi$
(which we will refer to as ``John'' term, following \cite{fab4}), gives a non-minimal kinetic coupling between the scalar field and gravity.
The theory without the Maxwell kinetic term and with $\beta\neq 0$ and $\gamma =0$ has been considered in \cite{Babichev:2013cya},
where various black hole solutions were presented, see also Refs. \cite{Bravo-Gaete:2013dca,Kobayashi:2014eva,Charmousis:2014zaa} for other solutions in arbitrary dimensions with different asymptotics. The last term in (\ref{action1}) mixes the scalar field and the gauge field.

Note that the action (\ref{action1}), although higher order in derivatives, gives second-order equations of motion, as we will explicitly see later.
The presence of the scalar field in (\ref{action1}) modifies the
dynamics of gravity. This scalar field is an ingredient for
modification of General Relativity, like in a generic Horndeski
model. The gauge field can be thought of as an extra modification, in which case it is unrelated to the electromagnetic field of the
standard model.

On the other hand, one can think of  $F_{\mu\nu}$ as of the standard
electromagnetic field. In this case, one should set $\gamma=0$ (or
$\gamma$ must be very small), otherwise photons would decay into the
scalar field, which would be detectable by various experiments. In
this context, the action (\ref{action1}) can be seen as a Horndeski
scalar-tensor theory plus matter, where the electromagnetic field
plays the role of matter. Note that in this case, certain electrically charged solutions have been reported in \cite{Cisterna:2014nua}.

In the first part of the paper we will
keep our discussion as general as possible, not specifying the
nature of the gauge field. In the second part we give particular
solutions, which usually corresponds to particular choices of
$\gamma$. Depending on this choice, a corresponding interpretation
of the gauge field should be given.

\subsection{Field equations and the ansatz}

Let us explicitly write down the equations of motion for the action
(\ref{action1}). The scalar field equation, obtained by varying the
action with respect to $\phi$, reads,
\begin{equation}
\nabla_{\mu}\left[  \left(\beta \,
G^{\mu\nu}-\eta\,g^{\mu\nu}-\gamma\,T^{\mu \nu}_{(M)}\right)
\nabla_{\nu}\phi\right]  =0. \label{eqphi}
\end{equation}
It is interesting to see that  (\ref{eqphi}) can be rewritten in
terms of a conserved current equation as
\begin{equation}
\nabla_\mu J^\mu =0,
\end{equation}
where the current $J^\mu$ is defined and given by
\begin{equation}\label{J}
    J^\mu:=J^\mu \equiv \frac{\delta \mathcal{L}}{\delta
\phi_{,\mu}}=\left(\beta \, G^{\mu\nu}-\eta\,g^{\mu\nu}-\gamma\,T^{\mu \nu}_{(M)}\right) \nabla_{\nu}\phi.
\end{equation}
This property is a consequence of the symmetry of the action
(\ref{action1}) with respect to the shift symmetry $\phi\to \phi+$const. The
quantity $J^\mu$ is naturally the Noether current associated with this
symmetry. Variation of (\ref{action1}) with respect to $A_\mu$ gives the gauge
field equation,
\begin{equation}
\partial_{\mu}\left\{\sqrt{-g}\left[F^{\mu\nu}-\gamma\,\left(\frac{1}{2}\,F^{\mu \nu}\,\nabla_{\sigma}\phi
+\big(F_{\sigma}^{\phantom{\mu} \mu}\nabla^{
\nu}\phi-F_{\sigma}^{\phantom{\mu} \nu}\,\nabla^{\mu}\phi
\big)\right)\,\nabla^{\sigma}\phi\right]\right\}=0\label{eqmaxwell}.
\end{equation}
And, finally, varying (\ref{action1}) with respect to the metric gives
the modified "Einstein" equations casted in the following form
\begin{eqnarray}
G_{\mu\nu}={T}_{\mu \nu}^{(1)}+{T}_{\mu \nu}^{(2)},\label{eqmetric}
\end{eqnarray}
where we defined,
\begin{eqnarray}\label{tmunu}
T_{\mu \nu}^{(1)}&=& -\Lambda\,g_{\mu \nu} + T_{\mu\nu}^{(M)},
\nonumber\\
T_{\mu\nu}^{(2)}  &
=&{\beta}\,\Big\{\frac{1}{2}\nabla_{\mu}\phi\nabla_{\nu }\phi
R-2\nabla_{\lambda}\phi\nabla_{(\mu}\phi
R_{\nu)}^{\lambda}-\nabla^{\lambda}\phi\nabla^{\rho}\phi
R_{\mu\lambda\nu\rho}
-(\nabla_{\mu}\nabla^{\lambda}\phi)(\nabla_{\nu}\nabla_{\lambda}%
\phi)+(\nabla_{\mu}\nabla_{\nu}\phi)\square\phi
\nonumber\\
&& +\frac{1}{2}G_{\mu\nu} (\nabla\phi)^{2} -g_{\mu\nu}\left[
-\frac{1}{2}(\nabla^{\lambda}\nabla^{\rho}\phi
)(\nabla_{\lambda}\nabla_{\rho}\phi)+\frac{1}{2}(\square\phi)^{2}%
-\nabla_{\lambda}\phi\nabla_{\rho}\phi R^{\lambda\rho}\right]\Big\}, \\
&+&\frac{1}{2}\,{\gamma}\,\left[F_{\mu\sigma}F_{\nu \rho}
\nabla^{\sigma }\phi\,\nabla^{\rho }\phi+\left(F_{\mu \sigma}
F^{\beta \sigma} \nabla_{\beta}\phi \nabla_{\nu}\phi+F_{\nu \sigma}
F^{\beta \sigma} \nabla_{\beta}\phi \nabla_{\mu}\phi\right)-\frac{1}{2}\, g_{\mu \nu}\,F_{\beta \sigma}F_{\tau}^{ \phantom{\tau} \sigma}\,\nabla^{\beta }\phi\,\nabla^{\tau }\phi\right.\nonumber\\
&&\left.+\frac{1}{8}\,g_{\mu \nu}\, \nabla^{\rho}\phi\,
\nabla_{\rho}\phi\, F_{\tau \beta} F^{\tau
\beta}-\frac{1}{2}\,F_{\mu \sigma} F_{\nu}^{\phantom{\nu}
\sigma}\,\nabla^{\rho}\phi\,\nabla_{\rho}\phi
-\frac{1}{4}\,\nabla_{\mu}\phi\,\nabla_{\nu}\phi\,F_{\tau \beta}
F^{\tau \beta}\right]\nonumber\\
&+& 2\eta\partial_{\mu}\phi\partial_{\nu}\phi-\eta g_{\mu\nu}\partial_{\sigma}\phi\partial^{\sigma}\phi.\nonumber
\end{eqnarray}
Note that $T_{\mu \nu}^{(1)}$ corresponds to the $\Lambda$-term and the
standard Maxwell field, as defined in (\ref{protoaction}), while
$T_{\mu \nu}^{(2)}$ comes from the the last three terms in
(\ref{action1}).

The above field equations look and are very complex. In the next subsection,
we, however, choose an ansatz, for which the equations of motion are
greatly simplified. Many components of the Einstein equations are
trivially zero due to the symmetries of the ansatz. The same is true
for the vector equation (\ref{eqmaxwell}). Moreover, as will be shown in a later section, we will establish a relation between one of
the components of the Noether current with one of the components of
the metric equations.

We assume the following static spherically symmetric ansatz for the
metric,
\begin{equation}\label{ag}
    ds^{2}=-h(r)\,dt^{2}+\frac{dr^{2}}{f(r)}+r^{2}\big(d \theta^{2}+\sin^{2}(\theta) d \varphi^{2}\big),
\end{equation}
while the scalar field is taken to be linearly time-dependent as
\begin{equation}\label{aphi}
    \phi(t,r)=q\,t+ \psi(r).
\end{equation}
Finally, we opt for a standard ``dyonic'' ansatz for the gauge field $A_\mu$,
\begin{eqnarray}\label{aA1}
    A_{\mu}dx^{\mu}=A(r)dt + B(\theta)d\varphi.
\end{eqnarray}
Note that as long as $B(\theta)$ is a general function of $\theta$, the
ansatz (\ref{aA1}) is not spherically symmetric. However, there is a
particular choice of $B(\theta)$, for which $A_{\mu}dx^{\mu}$ is
spherically symmetric. Indeed, the field strength two-form reads, $F
= \frac12 A'(r)dt \wedge dr + \frac12 B_\theta(\theta) d\theta\wedge
d\varphi$, where $B_\theta(\theta) \equiv dB(\theta)/d\theta$. The
first term of this expression only depends on the radial coordinate and
corresponds to the electric field. The second term is spherically
symmetric, only if $B_\theta(\theta) \propto \sin\theta$, because in
this case $B_\theta(\theta) d\theta\wedge d\varphi$ is proportional
to the element of the solid angle (i.e. the element of the area on a
unit sphere). In the Sec.~\ref{Secsol}, we will start from the ansatz
(\ref{aA1}), and we will see explicitly that the equations of motion require
$B(\theta)\propto \cos\theta$ reducing the ansatz (\ref{aA1}) to
\begin{eqnarray}\label{aA}
    A_{\mu}dx^{\mu}=A(r)dt - P\cos(\theta)d\varphi,
\end{eqnarray}
where $P$ is a constant.

\section{Solutions in an implicit form and their regularity}
\label{Secsol}

In this section we solve the set of field
equations~(\ref{eqphi}), (\ref{eqmaxwell}) and (\ref{eqmetric}),
assuming the ansatz~(\ref{ag}), (\ref{aphi}) and (\ref{aA1}). In the
first part of the Section, we give the general solution written in an implicit from. An auxiliary function of the
radial coordinate is introduced to simplify the problem, and this auxiliary function turns to be a solution of a fifth order algebraic
equation. The metric functions, the scalar field and the gauge field
can be deduced directly from this auxiliary function. For special choices of
the action parameters and the integration constants, it is possible
to reduce the algebraic equation on the auxiliary function to lower order and
write down the solution explicitly. In the phenomenologically
interesting case, $\gamma=0$, we present a perturbative solution
valid for small electric and magnetic charges.

\subsection{General solution in an implicit form}

We take the ansatz as in the previous section, i.e. in the
form~(\ref{ag}), (\ref{aphi}) and (\ref{aA1}). We will see in a
moment, that the equations of motion reduce (\ref{aA1}) to
(\ref{aA}). Indeed, defining
$$F(r)=A', \qquad C(\theta)=B_{\theta},$$
the Maxwell equations (\ref{eqmaxwell}) yield the following PDE,
\begin{equation}
\label{pardifeqmaxwell}
\begin{aligned}
 {\left\{\sqrt{\frac{f}{h}}\,r^{2}\, F\,
\left[1+\frac{\gamma}{2}\left(f\,(\psi')^{2}
-\frac{q^{2}}{h}\right)\right]\right\}'}
\left\{\frac{1}{r^{2}}\,{{\sqrt{\frac{h}{f}}}}
\left[1-\frac{\gamma}{2}\left(f\,(\psi')^{2}-\frac{q^{2}}{h}\right)\right]\right\}^{-1}\\
=\frac{1}{{\sin(\theta)}}\,{\left(\frac{C}{\sin(\theta)}\right)_{\theta}}.
\end{aligned}
\end{equation}
As we will see in  Sec. 5, the $(tr)$-component of the
Einstein equation is always satisfied if $J^r =0$, i.e.,
\begin{equation}\label{conJ}
    \beta\,G^{rr}-\eta\,g^{rr}-\gamma\,T^{rr}_{(M)}=0.
\end{equation}
Demanding then that (\ref{conJ}) is satisfied, the separability constant $C$ in (\ref{pardifeqmaxwell})
is found  to be zero. As a result, the vector potential
takes the form (\ref{aA}), as we expected. We note that, by satisfying
(\ref{conJ}), we solve simultaneously the $(tr)$-component of the
Einstein equation and the scalar field equation as we will see explicitly and in full generality in the forthcoming section. In other words the Anzatz (\ref{conJ}) that we impose is actually not only judicious but also necessary for a full solution to the field equations. At this point we can define our auxiliary function, setting
\begin{equation}\label{Sdef}
S \left( r \right) =\frac{\beta (r h(r))'  + \frac{\gamma}{4} r^2
F^2}{\eta r^2+\beta-\frac{\gamma P^2}{4 r^2} }
\end{equation}
so that  Eq.~(\ref{conJ}) reads, in terms of $S(r)$,
\begin{equation}\label{fSh}
f (r) ={\frac {h \left( r \right) }{S \left( r \right) }}.
\end{equation}
Our aim now is to give all unknown functions in terms of the function $S$. Given that (\ref{conJ}) guarantees simultaneously that the scalar field equation as well as the $(tr)$ equation are satisfied; hence it only remains to solve the $(r,r)$ and $(t,t)$ components of the
metric field equations. The $(r,r)-$component is an algebraic equation
on $\psi'$ yielding
\begin{equation}\label{xi}
\begin{aligned}
(\psi')^{2}&=\frac{1}{\left(4\,\eta\,{r}^{4}+4\,\beta\,{r}^{2}
-\gamma\,{P}^{2} \right)h(r)}\left\{ {\frac {{q}^{2} \left(
4\,\beta\,h' \left( r \right) +\gamma\, r  F \left( r \right)^{2}
\right) {r}^{3}}{ h \left( r \right) }}\right. \\
& +{\frac { \left( \gamma-\beta \right) {r}^{4} F \left( r
\right)^{2}}{ \beta }} +  \left. {\frac { \left[ {P}^{ 2} \left(
\gamma-\beta \right)-4\,{r}^{4} \left( \eta+\beta\,\Lambda \right)
\right] S \left( r \right) }{\beta}} \right\},
\end{aligned}
\end{equation}
while the $(t,t)$-component of the metric equation can be integrated once and written as
\begin{equation}\label{mastereqdyonic}
\begin{aligned}
\beta \left[{q}^{2}\beta-\frac{r^2}{4\beta}(\gamma-\beta)
F^2\right]-S(r) \left[ \left( \eta-\beta\,\Lambda \right)
{r}^{2}+2\,\beta -\frac{1}{4r^2}
P^2 (\beta+\gamma)\right]& \\
+C_{{0}} S(r)^{3/2} \left[\eta r^2+\beta-\frac{\gamma
P^2}{4r^2}\right]=0,&
\end{aligned}
\end{equation}
where $C_{0}$ is an integration constant.

The only remaining equation is the time component of the vector
equation (\ref{pardifeqmaxwell}), which reads, for our ansatz,
\begin{equation}\label{electric}
\sqrt{\frac{f}{h}}\,r^{2}\, F\,
\left[1+\frac{\gamma}{2}\left(f\,(\psi')^{2}
-\frac{q^{2}}{h}\right)\right]=Q,
\end{equation}
where $Q$ is a constant, related to the electric charge.

Using Eq.~(\ref{mastereqdyonic}) and its first derivative, the
electric equation (\ref{electric}) can be rewritten as,
\begin{eqnarray}
\left(\frac{\beta-\gamma}{S(r)^{1/2}}+\frac{ \gamma C_0}{2} \right)
\frac{F(r)}{\beta}=\frac{Q}{r^2}.\label{maxwelldyonic}
\end{eqnarray}
We can now treat (\ref{maxwelldyonic}) and (\ref{mastereqdyonic}) as two coupled algebraic equations with respect to $S$ and $F$ yielding a full solution to the system.

Equivalently we can simply substitute the above expression (\ref{maxwelldyonic}) into
(\ref{mastereqdyonic}) in order to get an algebraic equation of the fifth
order in $S^{1/2}(r)$ given by
\begin{eqnarray}  {r}^{2} \left( C_{{0}}\gamma\,\sqrt {S (r)
}+2\,\beta-2\, \gamma \right) ^{2} \,\left\{{\beta}^{2}{q}^{2}-
\left[ \left( \eta-\beta\,\Lambda \right) {r}^{2 }+2\,\beta-{\frac
{{P}^{2} \left(
\beta+\gamma \right) }{4 {r}^{2}} } \right] S(r)\right.&&\nonumber\\
\left.+C_0 S (r)^{3/2} \left( \eta\,{r}^{2}+\beta-{\frac
{\gamma\,{P}^{2 }}{4 {r}^{2}}} \right)\right\}- \left( \gamma-\beta
\right) {Q}^{2} S (r) {\beta}^{2}=0.&&
\end{eqnarray}
The system is therefore integrable in terms of a single algebraic equation on $S$ and the full ensemble of
solutions can be obtained, which will depend on three integration
constants $C_0$, $Q$ and $P$. The integration constant $C_0$
determines the asymptotic behavior of the solution. Indeed note
that (\ref{mastereqdyonic}) is independent of the mass of the black
hole solution since $S(r)$ depends on the derivative of $h(r)$. The
integration constant $Q$ is related to the electric charge, while
$P$ is the magnetic charge of a black hole. We have therefore reduced the full field equations to the solution of a single algebraic equation of the 5th order with respect to the auxiliary function $S$.

\subsection{Regularity for the implicit solutions}
We now argue, that a solution, given by the
equations (\ref{mastereqdyonic}) and (\ref{maxwelldyonic}),
generically contains a branch, which is regular at the future black
hole horizon. We will now show that if $r=r_H$ is an event horizon with $h(r_H)=0$ then the solution will be regular if $S(r)>0$ for all $r\geq r_H$.

Indeed, following \cite{Babichev:2013cya}, we
introduce the ingoing Eddington-Finkelstein coordinates,
\begin{equation}\label{v}
    v = t + \int ( fh )^{-1/2} dr.
\end{equation}
In the coordinates $(v,r)$, the metric takes the form,
\begin{equation}\label{metricEF}
    ds^2 = - h\,dv^2 +2\sqrt{h/f}\, dv\,dr + r^2 d\Omega^2.
\end{equation}
The ratio $h/f$ is simply $S$ from (\ref{fSh}), implying that the
metric is regular if $S(r)$ fulfills the above property.
Similarly, from (\ref{maxwelldyonic}) one can see that $F(r)$ is
regular, unless the expression in the parentheses is zero at the
horizon, which, however, requires some tuning of the parameters of
the model.

One may also worry about the regularity of the norm of the scalar
current, $J^2=g_{\mu\nu}J^\mu J^\nu$. Since the action is shift
symmetric, the norm of the current is a natural quantity to check.
For our solution, the radial component of the current is zero by
construction, see Eq.~(\ref{conJ}). Thus, the only non-zero component of
$J^\mu$ comes from the time component. From the definition of the
current, Eq.~(\ref{J}) and using (\ref{aphi}), we have,
\begin{equation}\label{J0}
    J^t = \left(\beta \, G^{tt}-\eta\,g^{tt}-\gamma\,T^{tt}_{(M)}\right) q.
\end{equation}
Note the presence of $g^{tt} =-1/h(r)$ in the above expression,
which is clearly divergent at the horizon and may be dangerous for regularity. However, other terms also
contain singular parts, which combine to a non-singular
expression. Indeed, calculating explicitly $G^{tt}$ for the ansatz
(\ref{ag}) and using (\ref{TE}), one finds,
\begin{equation}\label{Jt}
    J^t = \frac{1}{r^2 h^2}\left[ h\left( \eta r^2 +\beta -\frac{\gamma P^2}{4r^2} \right)-\beta h (fr)'-\frac14 \gamma r^2 f A'^2 \right] q,
\end{equation}
which, by virtue of (\ref{Sdef}), (\ref{fSh}) and
(\ref{Jt}) becomes,
\begin{equation}
J^t = \frac{\beta q}{r} \frac{S'(r)}{S(r)^{2}}.
\end{equation}
Again if $S(r)$ is smooth and non-zero at a horizon then the norm on the current
is not only finite, but it is in fact equal to zero,
\begin{equation}
\lim_{r\to r_H}J^2=0.
\end{equation}

It is interesting that not only the norm of the Noether current is
regular, but also the scalar field itself again under the same condition on $S$. Let us see how this works out. Near the horizon, $r\to
r_H$, $h(r_H)\to 0$,  and using (\ref{Sdef}), one can rewrite (\ref{xi})
as follows,
\begin{equation}\label{psir0}
    \psi'_{r=r_H} =\pm \frac{q}{h} S(r)^{1/2} + \mathcal{O}(1).
\end{equation}
This is of course singular and one would immediately conclude, if the scalar did not have a time dependance, that there is a scalar singularity there. However, substituting this expression  (with the plus sign) into (\ref{aphi})
and taking into account (\ref{v}), one finds,
\begin{equation}\label{phir0}
    \phi = qv -q \int dr\left( \frac{1}{\sqrt{fh}} - \frac{\sqrt{S(r)}}{h} \right) + \mathcal{O}(1).
\end{equation}
The expression inside the integral of (\ref{phir0}), naively, would
give a divergence of $\phi$ at the horizon, since it contains $f$
and $h$ in the denominator. However, using  (\ref{fSh}) one can
verify that the two terms inside the integral cancel each other,
thus yielding $\phi = qv+\mathcal{O}(1)$ at the horizon. Thus the
scalar field is not diverging at the future black hole horizon!

However, the scalar field, although regular at the future
horizon of the black hole, usually diverges at the de-Sitter
horizon, provided that the solution is asymptotically de-Sitter.
This happens because the coordinate $v$ is not regular at the future
de-Sitter horizon (independently of mass). Instead, the outgoing coordinate $u= t - \int (
fh )^{-1/2} dr$ is regular there{\footnote{We thank T. Kobayashi and N. Tanahashi for pointing this out to us.}}. The solution (\ref{psir0}) with
minus gives a regular solution for $\phi$ at the future de-Sitter
horizon. We should however warn  the reader that the action itself depends only on derivatives of the scalar field and not the scalar field itself and for the derivative field all quantities are finite and well-defined.

\section{Particular explicit solutions}

The master algebraic equation (\ref{mastereqdyonic}) is of fifth
order, and writing down a full solution involves complicated integrations which means that one can not write down the general solution
explicitly. Nevertheless, for some special cases the solutions can be written down
explicitly, as we will now see.

\subsection{Case $C_{0}=\Lambda=\eta=0$}
It is easy to note that the equations (\ref{mastereqdyonic}) and
(\ref{maxwelldyonic}) are considerably simplified for $C_0=0$, since
in this case (\ref{maxwelldyonic}) implies that $F(r)\propto
S(r)^{1/2}$, and  substituting  this into
(\ref{mastereqdyonic}), the latter becomes a linear algebraic
equation on $S(r)$. As a direct consequence $S(r)$ is found, and it is not difficult to
write down the solutions for all other functions. However, the full
expressions are still somewhat cumbersome; therefore, we present
here the solution for $\Lambda=\eta=0$.

The metric functions as well as scalar field profile take the
following form
\begin{equation}\label{C0}
\begin{aligned}
h(r) &= 1 - {\frac {\mu}{r}} - \frac{\sqrt{2}\left[
{P}^{2} \left(\gamma-\beta \right) - \left(\beta-3\,\gamma \right)
\bar{Q}^{2} \right]} {4\,r\, \sqrt{\beta\,[\bar{Q}^{2} \left( \gamma
- \beta \right) - \,{P}^{2}  \left( \beta+\gamma \right) ]}}
\arctan\left(\frac{2\,\sqrt{2\,\beta}\,r} {\sqrt{\bar{Q}^{2} \left( \gamma - \beta \right) - {P}^{2}  \left( \beta+\gamma \right) }}\right), \\
f(r) &=\left(1 + (\gamma-\beta)\frac{\bar{Q}^2}{ 8 r^2 \beta}-\frac{P^{2}\left(\gamma+\beta\right)}{8\,\beta\,r^{2}}\right)\,h(r), \\
(\psi'(r))^2 &=\frac{2}{\beta r^2 f(r)^2} \left(r^2\big(1-f(r)\big)+
\frac{\bar{Q}^2(\gamma-\beta)}{8
\beta}-\frac{P^{2}\left(\gamma+\beta\right)}{8\,\beta} \right),
\end{aligned}
\end{equation}
together with
\begin{equation}\label{C0bis}
F \left( r \right) =\pm \frac{\bar{Q} }{r^2}
\left(1+\frac{\bar{Q}^{2}(\gamma-\beta)}{8 r^2
\beta}-\frac{P^{2}\left(\beta+\gamma\right)}{8\,\beta\,r^{2}}\right)^{-1/2},
\;\; F_{\theta \varphi} =C(\theta)=P\,\sin(\theta),
\end{equation}
where  we defined the rescaled electric charge $\bar{Q}$ as,
$Q=\frac{\bar{Q} (\gamma-\beta)}{\beta}$. In the
above expressions we also fixed $q$ to have a particular value,
\begin{equation}\label{qbeta}
    q^2\beta = 2,
\end{equation}
such that $h(r)\to 1$ for asymptotically flat solutions, so the time
coordinate is defined properly and we don't have a solid deficit angle. In the above expressions we also implicitly assume
that
\begin{equation}\label{flatcon}
    \bar{Q}^{2} \left( \gamma - \beta \right) > \,{P}^{2}  \left( \beta+\gamma \right).
\end{equation}
In this case, the solution (\ref{C0}) and (\ref{C0bis}) corresponds
to an asymptotically flat black hole with $S$ strictly positive so that the scalar field is well defined beyond the event horizon of the black hole given by the condition $h(r_H)=0$. Both the metric and the
electromagnetic field are modified as compared to the charged black
hole solution in GR. Far from the black hole, that is for $r\to \infty$, the
electric field (\ref{C0bis}) has its usual form, $F\sim 1/r^2$,
however, with an effective charge $\bar Q$. On the other hand, the magnetic field is not modified. It is also
worthwhile to note that the mass of the black hole as measured at
infinity is not given simply by $\mu$, but it is modified, since the
last term in the first equation of (\ref{C0}) behaves as $1/r$ at
$r\to\infty$. 

One can also check that the norm of the Noether
current $J^2$ and the gauge field, (\ref{C0bis}) are regular at the
horizon. The scalar field (\ref{C0}) (for the branch with the plus
sign) is regular at the future black hole horizon.

When the condition (\ref{flatcon}) is not met, i.e. for $\bar{Q}^{2}
\left( \gamma - \beta \right) < \,{P}^{2}  \left( \beta+\gamma
\right)$, Eq.~(\ref{C0bis}) becomes imaginary at small $r$, thus the
solution is not physically viable and we do not discuss it.

By setting $P=0$, we easily recover the purely electric case, with no
magnetic charge. The condition for a well-defined asymptotically
flat solution is then simply $\gamma>\beta$. The purely magnetic
case is also easily obtained from the above solution by setting
$Q=0$. The parameter $\gamma$ in this case must have a negative
value, $\gamma<-\beta$.

Let us now consider another simple case for which the two metric functions are equal as it occurs for the standard Reissner-Nordstrom solution in GR.

\subsection{Case $f(r)=h(r)$}
The form of the expression for $S(r)$, see Eq.~(\ref{Sdef}) suggests
another explicitly solvable case. Indeed, assuming the two metric functions to be identical, $f(r)=h(r)$, then we
have
\begin{equation}\label{Sfh}
S(r)=1.
\end{equation}
Using (\ref{Sfh}), one easily finds that
\begin{equation}\label{solfh}
\begin{aligned}
h(r)&= 1-{\frac {\mu}{r}} +{\frac {\eta\,{r}^{2}}{3\,\beta}}+{\frac { \gamma\, \left( {Q}^{2}+{P}^{2} \right) }{4\,\beta\,{r}^{2}}}, \\
(\psi'(r))^2 &= \frac{1-f(r)}{f(r)^{2}} q^2,\\
F_{tr}& =F(r)=\frac{Q}{r^{2}}, \quad F_{\theta \varphi}
=C(\theta)=P\,\sin(\theta).
\end{aligned}
\end{equation}
The coupling constants, the constants of integration and the the
``velocity'' $q$ are related as
\begin{equation}\label{constfh}
{P}^{2}\beta\, \left( \Lambda\,\gamma+\eta \right) ={Q}^{2}\eta\,
 \left( \gamma-\beta \right)
,\qquad q^2=\frac{\eta + \Lambda\,\beta}{\beta\,\eta},\qquad
C_{0}=\frac{1}{\eta} \left(\eta-\beta\,\Lambda\right).
\end{equation}
The electromagnetic field is the same as in the case of the
Reissner-Nordstrom solution. The metric of this solution is
identical to the dyonic version of the Reissner-Nordstrom-de-Sitter
metric, however, up to the redefinitions of $Q$ and $P$. One can verify
that the metric, the gauge field and the norm of the Noether current
are regular at the horizon. The scalar field is regular at the
future black hole horizon for the branch with the plus sign, but
diverges at the future de-Sitter horizon.

When the magnetic charge is zero, $P=0$, the first equation of
(\ref{constfh}) implies  $\beta=\gamma$. On the other hand, when the
electric charge is zero, $Q=0$, then the same equation leads to the
relation $ \Lambda\,\gamma+\eta =0$, but in this case, the time-dependent character of the scalar field is lost.

This solution generalizes in a rather elegant way the self tuning black hole \cite{Babichev:2013cya} in the presence of an electric or magnetic charge. The self-tuning of the bulk cosmological constant is robust not only at the presence of gravitational mass but also in the presence of electro/magnetic charge.

Having found this stealth Reissner-Nordstrom-de-Sitter metric we can now find with relevant ease new solutions by a simple observation \cite{damos}.
Let us assume for simplicity that $P=0$ in which case this class of solutions dictates that $\beta=\gamma$. Then we have immediately that the electric strength has the usual Maxwell profile, $F(r)=\frac{Q}{r^{2}}$ and the equation for $S$ (\ref{mastereqdyonic}) is a third order polynomial equation in $\sqrt{S}$. Given that $\sqrt{S}-1$ is a root of this polynomial we simply divide the two, obtaining a second order polynomial equation for $\sqrt{S}$,
\begin{equation}
\label{damos}
S-\frac{u}{R^2+1}S^{1/2}-\frac{u}{R^2+1}=0
\end{equation}
where $$u=\frac{\eta+\beta \Lambda}{\eta-\beta \Lambda}$$ and we have rescaled the radial coordinate to $R=\sqrt{\frac{\eta}{\beta}}r$. It is then a straightforward exercise to read off the charged black hole solutions corresponding to the two remaining roots,
\begin{equation}
h(R)=u-\frac{\mu}{R}+\frac{\eta Q^2}{4\beta\,R^2}+\frac{u^2}{2R}\left[\arctan{R}\pm \arctan{\frac{R\sqrt{u}}{\sqrt{4R^2+4+u}}}\right]\pm \frac{u^{3/2}}{R}\sinh^{-1}\frac{2R}{\sqrt{u+4}}
\end{equation}
with $f(R)=\frac{h(R)}{S(R)}$ where $S$ is the square of the solution(s) of (\ref{damos}). The scalar field can be directly calculated from (\ref{xi}) and is regular.
This solution is the charged version of the black hole solution found in \cite{damos}. Note that for $\Lambda=0$ we have $u=1$, and we have well defined asymptotics for $h$. Note that for large R we have,
$$
h(R)\sim 1-\frac{C_1 \ln R }{R}-\frac{C_2}{R}+\frac{C_3}{R^2}+O(1/R^3)
$$
with $C_i=C_i(u,\mu, Q)$; in other words we have no longer a standard Newtonian fall-off  and the mass and charge terms are rescaled. Also we note that $f(R)\sim R^2$, and hence our spacetime asymptotes an Einstein static universe.

\subsection{Perturbative solution for $\gamma=0$ and $\eta=\Lambda=0$.}

In this subsection we consider the phenomenologically interesting case
$\gamma=0$ where the electric field is a Maxwell field which is not coupled to the scalar Galileon field. We could not find simple enough explicit solutions to
Eqs.~(\ref{mastereqdyonic}), (\ref{maxwelldyonic}). Instead, we
perform a perturbative analysis for small charges $Q$ and $P$. For
$P=Q=0$ and $\gamma=0$, Eq.~(\ref{maxwelldyonic}) is satisfied for
$F=0$, that is the gauge field is identically zero.
Eq.~(\ref{mastereqdyonic}) is solved then for $S=S_0$, where $S_0$
is a constant, which depends on $C_0$, $\beta$ and $q$. Substituting
$S=S_0$ into (\ref{Sdef}) we find, $h_0(r) = S_0-\frac{\mu}r$, where
$\mu$ is a constant of integration. To have canonically defined time
at spatial infinity we have to set $S_0 =1$, which fixes $C_0$ from
(\ref{mastereqdyonic}) as follows, $C_0=2-q^2\beta$.
From~(\ref{fSh}) one can easily see that
\begin{equation}\label{pertfh0}
    f_0(r)=h_0(r) = 1-\frac{\mu}{r},
\end{equation}
and from (\ref{xi}) we have,
\begin{equation}\label{pertpsi0}
    \psi'_0=\pm \frac{q}{h_0}\sqrt{rh'_0}.
\end{equation}
This solution is nothing but the stealth solution defined on the Schwarzschild metric found in \cite{Babichev:2013cya}.
Now we assume that the charges $Q$ and $P$ are non-zero, but "small",
and we find a perturbative expansion in $Q$ and $P$ around the
solution (\ref{pertfh0}), (\ref{pertpsi0}).
Eq.~(\ref{maxwelldyonic}), after the redefinition $Q\to Q\beta^2$,
implies the standard Maxwell law,
\begin{equation}\label{pertF}
    F= \frac{Q}{r^2}.
\end{equation}
Substituting (\ref{pertF}) into (\ref{mastereqdyonic}) one gets a
correction to $S(r)$,
\begin{equation}\label{pertS}
    S(r) = 1 -\left(\frac{1}{2 r^2}\right)\frac{Q^2 + P^2}{3q^2\beta -2}.
\end{equation}
linear in $Q^2 + P^2$ the strength of the EM field.
It is not difficult then to find the leading order corrections to
the metric,
\begin{equation}\label{pertfh}
    \begin{aligned}
    h  &= 1- \frac{\mu}{r} + \frac{Q^2+P^2}{2(3q^2\beta -2)}\frac{1}{r^2}, \\
    f & = h(r)\left(1 +\frac{Q^2 + P^2}{2 r^2(3q^2\beta -2)}\right),
    \end{aligned}
\end{equation}
and  to the scalar field,
\begin{equation}\label{pertpsi}
    \psi' = \pm \frac{ q \sqrt{rh'}}{h}\left( 1- \frac{Q^2+P^2}{8\beta q^2 r^3 h'} \right).
\end{equation}
The metric (\ref{pertfh}) is asymptotically flat. It is not of the
Reissner-Nordstrom form, since $h(r)\neq f(r)$ at the leading order
in $Q$ and $P$. The solution (\ref{pertF}), (\ref{pertfh}) and
(\ref{pertpsi}) is valid as long as the charges $Q$ and $P$ satisfy the relation,
\begin{equation}
    Q^2 + P^2 \ll 2 r^2 |(3q^2\beta -2)|.
\end{equation}
which is true generically before we hit the event horizon of the black hole.
Again, one can easily check that the solution (\ref{pertF}),
(\ref{pertfh}) and (\ref{pertpsi}) with the plus sign is regular at
the future black hole horizon.

\subsection{Static case for $\beta=\gamma$}
\label{Staticcase}
Let us now consider the case $q=0$. The scalar field then has no time dependence and as a result we expect for the scalar field to be singular at the horizon. However, these solutions are not immediately rejected because the action depends only scalar derivatives and they are in fact regular. So we  study briefly this case here.
Previously, progress in this direction was reported in \cite{Cisterna:2014nua} where the authors considered the static $q=0$ case but for a minimally coupled gauge field $\gamma=0$. Here we will consider the case $\beta=\gamma$.
The auxiliary function $S(r)$ takes the form,
\begin{equation}
S(r)={\frac {4 \left[  2\left( \Lambda\,\beta-\eta \right) {r}^{4}-4
\,\beta\,{r}^{2}+{P}^{2}\beta \right] ^{2}}{{C_{{0}}}^{2}\left(4\,
\eta\, r^{4}+4\, \beta\, r^{2}-P^{2}\,\beta\right)^{2}}},
\end{equation}
For simplicity we will consider only the electric case and set $P=0$ before evaluating the relevant functions. The dyonic or magnetic solution are obtained with ease using the above $S$.

For $P=0$ we have,
\begin{eqnarray}
h(r)&=&{\frac { \left( \eta-\beta\,\Lambda \right)
^{2}{r}^{2}}{3\,\eta\,{C _{{0}}}^{2}}}+{\frac {\left(
3\,\eta+\beta\,\Lambda \right)
 \left( \eta-\beta\,\Lambda \right) }{{C_{{0}}}^{2}{\eta}^{2}}}+\frac{{\beta
} \left( \eta+\beta\,\Lambda \right) ^{2}}{{{\sqrt
{\eta\,\beta}{C_{{0}}}^{2}{\eta}^{2}{r}} }}\arctan \left( {\frac {
\eta\,r}{\sqrt {\eta\,\beta}}} \right)+{\frac
{{Q}^{2}}{{r}^{2}}}-\frac{\mu}{r},
\nonumber\\
f(r)&=&{\frac {{C_{{0}}}^{2} \left( \eta\,{r}^{2}+\beta \right)
^{2}}{
 \left[ (\eta-\beta\,\Lambda)\,r^{2}+2\,\beta \right] ^{2}\beta
}} \,h(r),
\nonumber\\
(\psi'(r))^2&=& -{\frac {{r}^{2} \left( \eta+\beta\,\Lambda \right)
}{\beta\, \left( \eta\,{r}^{2}+\beta \right) \,f(r)}},
\end{eqnarray}
Note here that the scalar function $\psi(r)$ will be singular since there is no time dependence to cancel out the radial singularity. However the norm $(\nabla \phi)^2$ appearing in the action is regular and finite at the horizon.
The electric field strength is given by,
\begin{equation}
F_{rt}=B(r)=\frac{2\,Q}{r^{2}}.
\end{equation}
and it is noteworthy that the field is identical to a Maxwell field for spherical asymmetry unlike what happens in the $\gamma=0$ case.

\section{Relation between the $(tr)$-component of the Einstein equation and the Noether current.}
As we mentioned in the introduction we would now like to take a step back and prove that our ansatz (\ref{ag}), (\ref{aphi}) and
(\ref{aA}) is mathematically consistent. This is not trivial since the scalar field does not obey the same symmetries that the spacetime. It turns out that there is an interesting relation between the $(tr)$-component of the metric equation associated with the flux of energy of the scalar field and the
Noether current associated with the shift symmetry of the action.
In this section our hypothesis is the above ansatz but we do not require a specific form of the
action~(\ref{action1}), but rather consider a general action given by
\begin{equation}\label{act2}
    S = \int d^4 x\, \mathcal{L}(g_{\mu\nu}, \nabla\phi, \nabla^2\phi, F_{\mu\nu}),
\end{equation}
In the above we assume that the action does not depend on
$\phi$ explicitly, but only on its derivatives (as is the case for (\ref{action1})); this requirement ensures the existence of a Noether conserved current central to our whole
discussion\footnote{The expression for Noether current in the Horndeski theory can be found in~\cite{Gao:2011qe}.}.
We also require that the vector field is of Maxwell type,
i.e. it enters only via the gauge invariant combination $F_{\mu\nu}
\equiv \partial_\mu A_\nu - \partial_\nu A_\mu$. These properties
are crucial for finding black hole solutions with non-trivial scalar
configurations as we saw earlier on. Let us analyse the structure
of the equations of motion for the ansatz (\ref{ag}), (\ref{aphi}).

The equations of motion for the action~(\ref{act2}) can be written
generically as,
\begin{eqnarray}
    \mathcal{E}^{(\phi)} &\equiv & \frac{1}{\sqrt{-g}}\frac{\delta \mathcal{L}}{\delta \phi} =0, \label{eomphi}\\
    \mathcal{E}_{(A)}^{\mu} &\equiv&  \frac{1}{\sqrt{-g}}\frac{\delta \mathcal{L}}{\delta A_\mu} =0, \label{eomA}\\
     \mathcal{E}^{(g)}_{\mu\nu} & \equiv& \frac{2}{\sqrt{-g}} \frac{\delta\mathcal{L}}{\delta g^{\mu\nu}} =0. \label{eomg}
\end{eqnarray}

For the ansatz~(\ref{ag}), (\ref{aphi}) and (\ref{aA}) the
non-trivial components of the metric equation are the $(tt)$,
$(tr)$, $(rr)$ and $(\theta\theta)$ components (the
$(\varphi\varphi)$ component is also non-trivial, but it is a simple
consequence of the $(\theta\theta)$ component, due to spherical
symmetry,
$\mathcal{E}^{(g)}_{\theta\theta}\sin^2\theta=\mathcal{E}^{(g)}_{\varphi\varphi}$).
Thus, there are four metric equations. There is one scalar equation,
$\mathcal{E}^{(\phi)}=0$ and one non-trivial
component of the gauge field equation, the time component. Thus, there
are in total six equations. Now notice that the ansatz~(\ref{ag}),
(\ref{aphi}) and (\ref{aA}) contains only four independent
functions, namely, $f(r)$, $h(r)$, $\phi(r)$ and $A(r)$. This {\it a
priori} means that the system of equations is overdetermined.
However, as we will now demonstrate, not all the equations are independent.

Let us consider the infinitesimal coordinate transformation,
\begin{equation}\label{coord}
    x^\mu \to \tilde{x}^\mu = x^\mu + \xi^\mu.
\end{equation}
The scalar field in the new coordinates is $   \tilde{\phi}(\tilde{x})
= \phi(x)$ which, expanded in terms of $x^\mu$, gives, $\tilde{\phi}(\tilde{x}) =  \tilde{\phi}(x)+
\phi_{,\mu}\xi^\mu+\mathcal{O}(\xi^2)$, yielding
\begin{equation}
    \tilde{\phi}(x) =  \phi(x) - \phi_{,\mu}\xi^\mu.
\end{equation}
to leading order.
Similarly, for the metric we have, $
    \tilde{g}^{\mu\nu} (\tilde{x}) =  g^{\mu\nu} (x) + g^{\alpha\mu}\xi^\nu_{,\alpha} + g^{\alpha\nu}\xi^\mu_{,\alpha},
$ and, on the other hand, $
    \tilde{g}^{\mu\nu} (\tilde{x}) = \tilde{g}^{\mu\nu} (x) +\partial_\beta g^{\mu\nu}\xi^{\beta},
$ which combine to,
\begin{equation}
    \tilde{g}^{\mu\nu} (x) = g^{\mu\nu} (x)+  g^{\alpha\mu}\xi^\nu_{,\alpha} + g^{\alpha\nu}\xi^\mu_{,\alpha} -\partial_\beta g^{\mu\nu}\xi^{\beta}
        = g^{\mu\nu}(x) + \xi^{\mu;\nu}+ \xi^{\nu;\mu}.
\end{equation}
Under the infinitesimal change of coordinates (\ref{coord}) the
vector field transforms as,
\begin{equation}
    \tilde{A}_{\mu} (x) = A_\mu(x) - ( \xi^{\alpha}A_\alpha)_{,\mu}.
\end{equation}
Since the action~(\ref{act2}) is invariant under the coordinate
changes, we obtain,
\begin{equation}\label{mas}
    \begin{aligned}
    \delta S
    =- \int d^4x  \sqrt{-g} \, \mathcal{E}^{(\phi)} \, \phi{,_\nu} \xi^\nu - \int d^4 x\,\sqrt{-g} \, \mathcal{E}_{(A)}^{\mu}  ( \xi^{\alpha}A_\alpha)_{,\mu}
    + \int d^4x \sqrt{-g} \, \mathcal{E}^{(g)}_{\mu\nu} \xi^{\mu;\nu} =0.
    \end{aligned}
\end{equation}
Integrating by parts one can get rid of the second term since
\begin{equation}
\nabla_\mu\mathcal{E}_{(A)}^{\mu} =
\frac1{\sqrt{-g}}\partial_t(\sqrt{-g} \mathcal{E}_{(A)}^t) = 0.
\label{A0}
\end{equation}
Taking this into account and integrating by parts the last integral
(\ref{mas}), we obtain,
\begin{equation}\label{relate}
    \mathcal{E}^{(\phi)}\phi_{,\nu}   + \nabla^\mu \mathcal{E}^{(g)}_{\mu\nu} =0.
\end{equation}
for arbitrary $\xi^\nu$.
Now, Eq.~(\ref{relate}) consists of two non-trivial relations, the time and
radial components,
\begin{eqnarray}
    q \mathcal{E}^{(\phi)} + \frac1{\sqrt{-g}}\partial_r \left( \sqrt{-g} g^{rr}\mathcal{E}_{tr}^{(g)}\right) &=&0, \label{relatet}\\
    \mathcal{E}^{(\phi)}\psi'(r) + \frac1{\sqrt{-g}}\partial_\mu \left( \sqrt{-g} g^{\mu\nu}\mathcal{E}_{\nu r}^{(g)}\right)
    -\frac12\frac{\partial g_{\mu\nu}}{\partial r} \mathcal{E}^{\mu\nu}_{(g)}&=& 0.
    \label{relater}
\end{eqnarray}
As we can see, two among the six  non-trivial equations of motion are not
independent. E.g., using (\ref{relatet}), one can express the scalar
equation in terms of the $(tr)$-component of the Einstein equation.
Then, the $(\theta\theta)$-component can be found from
(\ref{relater}) in terms of $(tr)$, $(tt)$ and $(rr)$ components of
the Einstein equation. We have four independent ordinary differential
equations with four functions, hence the system of equations is mathematically consistent.

This is sufficient but we can obtain an integral version of (\ref{relatet}),
which nicely connects the Noether current associated with the shift
symmetry $\phi\to\phi +$const with the scalar flux $\mathcal{E}_{tr}^{(g)}$. Indeed,
since the action~(\ref{act2}) does not explicitly depend on $\phi$,
but only on its derivatives, the equation of motion for the
scalar field can be written as $\mathcal{E}^{(\phi)} = \nabla_\mu J^\mu
=0$, where $J^\mu \equiv \frac{\delta \mathcal{L}}{\delta
\phi_{,\mu}}$ is the Noether current. Therefore, substituting
$\mathcal{E}^{(\phi)}=\nabla_\mu J^\mu$ into (\ref{relatet}) and
integrating over the radial coordinate, one finds,
\begin{equation}
        q J^r +  g^{rr}\mathcal{E}_{tr}^{(g)} =\text{const.} \label{relateI}\\
\end{equation}
The constant in (\ref{relateI}) is yet to be determined. It is not
arbitrary and in fact,  a more careful investigation shows that it is
zero. Indeed, from (\ref{mas}) we obtain, integrating by parts
(and taking into account (\ref{A0})),
\begin{equation}\label{mas2}
     \int d^4x  \sqrt{-g} \,  J^\mu \, \nabla_\mu \left( \phi{,_\nu} \xi^\nu\right) + \int d^4x \sqrt{-g} \,  \mathcal{E}^{(g)}_{\mu\nu} \xi^{\mu;\nu} =0.
\end{equation}
To flesh out the result let us consider a particular infinitesimal change of
coordinates, see~(\ref{coord}),
$$\xi^\mu = (\xi(r),0,0,0).$$
Taking into account that for the assumptions above the only non-zero
component of $\xi^{\mu;\nu}$ are $\xi^{t;r}$ and $\xi^{r;t}$, Eq.
(\ref{mas2}) can be rewritten as,
\begin{equation}
    \label{mas3}
     \int d^4x  \sqrt{-g} \,  J^r \, \partial_r \left( q \xi(r) \right) + \int d^4x \sqrt{-g} \,  \mathcal{E}^{(g)}_{tr} \left( \xi^{t;r}(r) +  \xi^{r;t}(r)\right)  =0.
\end{equation}
Now, remarkably, for the ansatz above, one finds $\xi^{t;r}(r) +
\xi^{r;t}(r) = g^{rr} \xi_{,r}(r)$. Thus from the last equation we
obtain,
\begin{equation}\label{TJ}
    -q J^r = \mathcal{E}^{(g)}_{tr} g^{rr}.
\end{equation}
The above equation (\ref{TJ}) is identical to (\ref{relateI}) for
const=0. We have thus demonstrated that our time dependent scalar field Anzatz is in fact compatible with the absence of flux for the scalar field and thus a general static and spherically symmetric metric.

\section{Discussion}
\label{SecDisc}

In this paper we discovered new exact black hole solutions
in a scalar-tensor-vector theory, given by~(\ref{action1}). The action, besides the Einstein-Hilbert
term, consists of the standard kinetic term for the scalar field,
the standard kinetic Maxwell term for the gauge field, and also two
non-linear terms, describing a kinetic (non-minimal) mixing of the
scalar field with the graviton, and a kinetic mixing of the scalar
field with the gauge field. The action is shift-symmetric in the
scalar field, i.e. the action depends on the derivatives of the
scalar field, and not on $\phi$ itself. Also, the vector field
enters the action only through the Maxwell combination
$(\partial_\mu A_\nu-\partial_\nu A_\mu)$, therefore the action is
gauge invariant.

We searched solutions for a static and spherically
symmetric ansatz for the metric (\ref{ag}) and the gauge field
(\ref{aA}), while for the scalar field we crucially assumed a time (and radial) dependent
ansatz of the form (\ref{aphi}). We showed that this ansatz is consistent as long as the time dependence of the scalar is linear. The  equations of motion (\ref{eqphi}), (\ref{eqmaxwell}) and
(\ref{eqmetric}) reduce to four independent ODEs with four functions
entering the ansatz. We also we were able to prove that for this
particular ansatz the $r$-component of the Noether current,
associated with the shift symmetry, is proportional to the
$(tr)$-component of the Einstein equations, see Eq.~(\ref{TJ}).

The system of the field equations for the precise ansatz we consider turns out to be
integrable: we obtained two algebraic equations
(\ref{mastereqdyonic}) and (\ref{maxwelldyonic}), which fully
determine the solution. Having found the auxiliary function $S(r)$
and the electric field, $F(r)$ from this algebraic equations, other
functions of the ansatz can be calculated from (\ref{Sdef}),
(\ref{fSh}) and (\ref{xi}). The solution contains three integration
constants: $Q$ is proportional to the electric charge of the black
hole, $P$ is related to the magnetic charge of the monopole, and the
constant $C_0$ which
determines the asymptotic behavior of the solution. Note that there is a fourth
integration constant (related to the mass of the black hole), which is not
present in the master equations (\ref{mastereqdyonic}) and
(\ref{maxwelldyonic}), however, it appear when one integrates
(\ref{Sdef}) for a given $S(r)$ to find $h(r)$. Depending on the
choice of parameters and the integration constants, the solution
describes a black hole with various asymptotic behaviors at
infinity. An important point is that for a generic choice of
parameters the metric, the gauge field  and the norm of the scalar
current are regular at the horizon(s). As for the scalar field
itself, its solution normally contains a branch, which is regular at
the future horizon. If the metric is asymptotically de-Sitter, the scalar
field is regular either i) at the future black hole horizon and at
the past de-Sitter horizon, or ii) at the future de-Sitter horizon
and the past black hole horizon.

For some choices of the parameters of the action and the constants
of integration we gave explicit solutions, see Sec.~\ref{Secsol}. In particular, we studied a solution for
$\gamma=0$. This case is interesting from the phenomenological point
of view, since it corresponds to a minimal coupling of the gauge
field to the gravity, therefore it makes sense to think of the gauge
field as the electromagnetic field of the Standard model. The action
(\ref{action1}) in such a case describes
Horndeski theory with a matter field.

Here, we have presented a quite detailed scan of the analytic solutions and it will be interesting to explore the thermodynamics of the configurations
found in this paper and to compare them with the standard Reissner-Nordstrom solution for example (which corresponds to the case $\beta=\eta=\gamma=0$). This issue is rendered non trivial by the fact that the scalar field is time-dependent, and hence as a first step in this task, one can analyze the purely static case, see Section (\ref{Staticcase}).

One can also explore if there exists more general Horndeski sectors with a shift symmetry that may be integrated as in the present case. For spherical symmetry for example it is clear that our result in the last section is a very important step in this direction. To simplify the discussion let us assume no gauge field but a general shift symmetric theory. For the anzatz of the scalar field that we consider taking $J^r=0$ renders the system in fact integrable in the following sense. Indeed one is left with the $(r,r)$ and $(t,t)$ equations to solve with respect to the radial scalar $\psi'(r)$ and the metric function $h(r)$. The $(r,r)$ is an algebraic equation with respect to $\psi'(r)$ and hence can be in principle solved algebraically. Then, at the end, we are left with a non-linear ODE with respect to $h(r)$ which can be solved analytically or numerically.

Another important aspect is to study cases of lesser symmetry, for example not static but rather stationary solutions. There, one would like to go further and find an argument, such as the one for the radial part of the current being zero, which again will render the field equations integrable. We hope to report progress on these topics in the near future.

{\it Acknowledgments}. The work of E.B. was supported in part by the Grants No. RFBR 13-02-00257, 15-02-05038. The work of M.H. is partially supported by grant
1130423 from FONDECYT and from CONICYT, Departamento de Relaciones Internacionales ``Programa Regional MATHAMSUD 13 MATH-05'', and M. H. thanks Moises Bravo for useful discussions.



\end{document}